\documentclass[prl,twocolumn,letterpaper, superscriptaddress]{revtex4-1}
\usepackage{graphicx}
\usepackage{dcolumn}
\usepackage{bm}
\usepackage{color}
\usepackage{tabularx}
\usepackage{array}
\usepackage{amsmath}
\usepackage{pbox}
\usepackage{stmaryrd}  

\begin{document}

\title{On the Temperature-dependent \textcolor{black}{Characteristics} of \textcolor{black}{Perpendicular Shape Anisotropy-Spin Transfer Torque-Magnetic Random Access Memories (PSA-STT-MRAMs)}}

\author{Wei Zhang}
\thanks{Correspondence to: W. Zhang, Email: weizhang@oakland.edu}
\affiliation{Department of Physics, Oakland University, Rochester, MI 48309, USA}

\author{Zihan Tong}
\affiliation{Department of Electronic and Computer Engineering, The Hong Kong University of Science and Technology, Clear Water Bay, Kowloon, Hong Kong, China}

\author{Yuzan Xiong}
\affiliation{Department of Physics, Oakland University, Rochester, MI 48309, USA}

\author{Weigang Wang}
\affiliation{Department of Physics, University of Arizona, Tucson, AZ 85721, USA}

\author{Qiming Shao}
\affiliation{Department of Electronic and Computer Engineering, The Hong Kong University of Science and Technology, Clear Water Bay, Kowloon, Hong Kong, China}
\affiliation{Department of Physics, The Hong Kong University of Science and Technology, Clear Water Bay, Kowloon, Hong Kong, China}

\date{\today}

\begin{abstract}

The \textcolor{black}{perpendicular shape anisotropy-spin transfer torque-magnetic random access memories (PSA-STT-MRAMs)} takes advantage of the nanopillar free-layer geometry for securing a good thermal stability factor from the shape anisotropy of the nanomagnet. Such a concept is particularly well-suited for small junctions down to a few nanometers. At such a volume size, the nanopillar can be effectively modeled as a Stoner-Wohlfarth (SW) particle, and the shape anisotropy scales with the spontaneous magnetization by $\sim M^2_s$. For almost all ferromagnets, $M_s$ is a strong function of temperature, therefore, the temperature-dependent shape anisotropy is an important factor to be considered in any modeling of the temperature-dependent performance of \textcolor{black}{PSA-STT-MRAMs}. In this work, we summarize and discuss various possible temperature-dependent contributions to the thermal stability factor and coercivity of the \textcolor{black}{PSA-STT-MRAMs} by modeling and comparing different temperature scaling and parameters. We reveal nontrivial corrections to the thermal stability factor by considering \textcolor{black}{both} temperature-dependent shape and interfacial anisotropies. The coercivity, blocking temperature, and electrical switching characteristics that resulted from incorporating such a temperature dependence are also discussed, in conjugation with the nanomagnet dimension and coherence volume. 

\end{abstract}

\maketitle

\section{Introduction}

\textcolor{black}{Magnetic tunnel junctions (MTJs) have been intensively investigated fundamentally and developed technologically in the past decades due to their high potential in future non-volatile magnetic memories.} While the end of the Moore's law may be approaching, the scaling of the MTJs has also been one of the main challenges in the development of nonvolatile spin-transfer torque magnetoresistive random access memory (STT-MRAM). The key technology of today's MTJ is the perpendicular MTJ (p-MTJ) using interfacial anisotropy at a ferromagnet/oxide (e.g. FeCoB/MgO) interface, and high volume manufacturing based on p-MTJs with only tens of nm has been initiated. However, as MTJ units become smaller to meet the ultrahigh-density integration, their thermal stability factor becomes a critical issue, as the interfacial anisotropy inevitably reaches a physical limit in securing sufficient energy barrier between the two possible states. 

\begin{figure*}[htb]
 \centering
 \includegraphics[width=6.7 in]{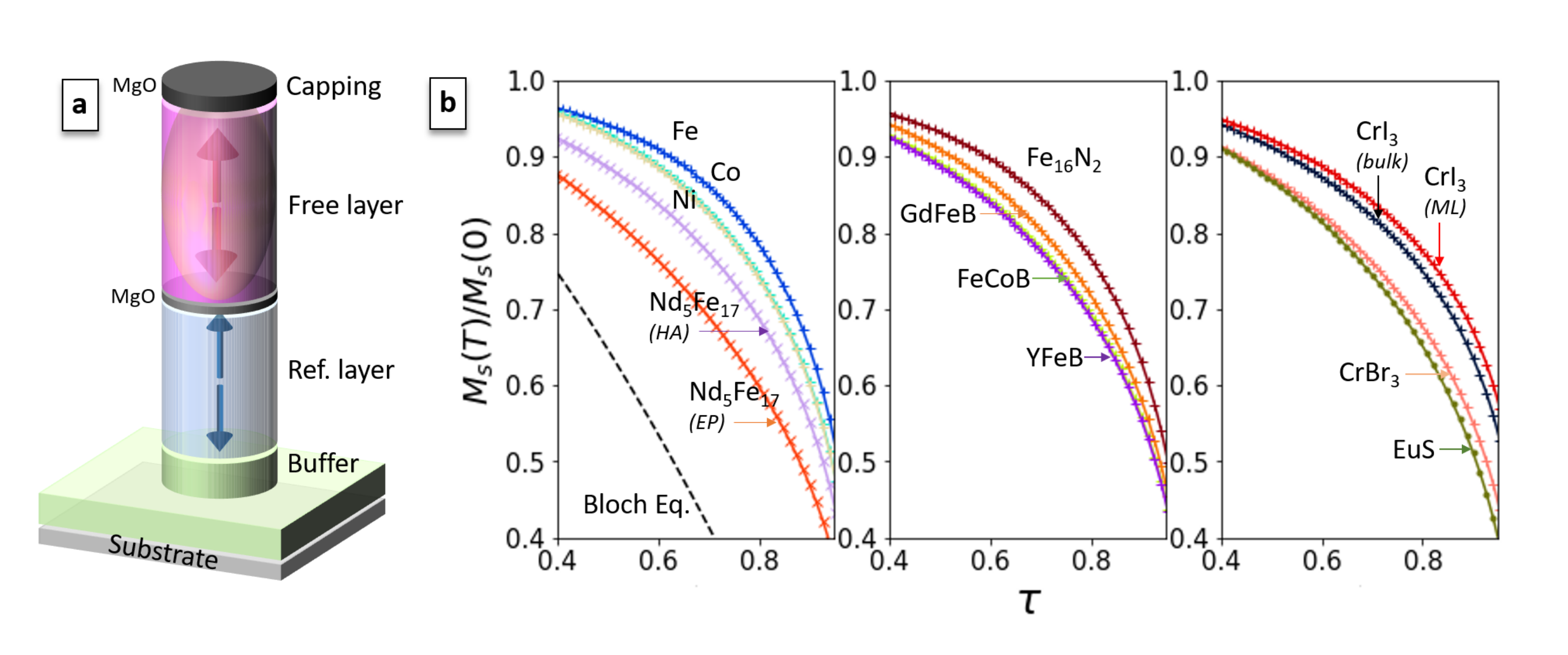}
 \caption{(a) Schematic illustration of the \textcolor{black}{PSA-STT-MRAM}, consisting (from the bottom): buffer layer, reference layer, MgO layer (for inducing $K_i$), free-layer, MgO and capping layer. Compared with conventional MTJ, the free-layer magnet in this \textcolor{black}{PSA}-MTJ is made into a nanopillar with $t>>D$ so that the easy axis is along the thickness direction. (b) Temperature scaling of $m(\tau)$: the Bloch model for the temperature dependence (dashed line) and the re-plotted fitting curves (symbols) adopted from previous experimental reports using the Kuz'min temperature scaling with fitting parameters ($s, p$) for example materials, Fe (0.35, 4), Co (0.11, 2.5), Ni (0.15, 2.5) \cite{kuzmin1,kuzmin2}, Nd$_5$Fe$_{17}$ (EP: 1.49, 2,5. HA: 0.72, 2.5) \cite{yu_jac2018}, GdFeB (0.4, 2.5) \cite{kuzmin_jap2010}, FeCoB (0.65, 2.5) \cite{jian_ieee2014}, YFeB (0.7, 2.5) \cite{kuzmin_jap2010}, Fe$_{16}$N$_2$ (0.42, 3.8) \cite{dirba_acta2017}, CrI$_3$ (bulk: 0, 1.7, Kuz'min scaling: 0.25. ML: 0, 1.7, Kuz'min scaling: 0.22) \cite{wahab_am2021}, CrBr$_3$ (1, 2.5, Kuz'min scaling: 0.31), and EuS (0.8, 2.5, Kuz'min scaling: 0.369) \cite{kuzmin_pla2005}. }
 \label{fig1}
\end{figure*}

As a result, a new concept known as the \textcolor{black}{perpendicular shape anisotropy-spin transfer torque-magnetic random access memories (PSA-STT-MRAMs)} has been recently proposed and demonstrated \cite{watanabe_ncomm2018,dieny_nanoscale2018,dieny_jphysd2019,jinnai_apl2020,jinnai_iedm2020}, which synergizes the shape anisotropy of the free-layer nanomagnet with the interfacial anisotropy to achieve high perpendicular anisotropy by properly engineering the free-layer's aspect-ratio. This concept is well suited for future scaling of MTJs in the context of using nanomagnets for device building blocks, as the free-layer magnet approaching a \textcolor{black}{Stoner}-Wohlfarth (SW) particle with dimensions of a few tens of nm$^3$ or even smaller \cite{jinnai_iedm2020}, as illustrated in Fig. \ref{fig1}(a).   
 
In \textcolor{black}{PSA-STT-MRAMs}, assuming the single-domain magnetization reversal of the free-layer, the thermal stability factor, $\Delta$, is expressed as \cite{watanabe_ncomm2018}: 

\begin{equation}
\Delta = \frac{E_0}{k_B T} = (-\delta N \frac{M_{s0}^2}{2 \mu_0} t + K_b t + K_i) \frac{\pi D^2}{4 k_B T}, 
\label{eq01}
\end{equation} 
where $\mu_0$ is the permeability in free space, \textcolor{black}{$E_0$} the energy barrier, $k_B$ the Boltzmann constant, $M_{s0}$ the saturation magnetization of the bulk ferromagnet, and $T$ the absolute temperature. Energy-wise, $K_b$ and $K_i$ are the bulk (magneto-crystalline) and interfacial-anisotropy energy densities, respectively. In conventional MTJs, the primary perpendicular anisotropy contribution comes from the interfacial term, $K_i$. Dimension-wise, $t$ and $D$ are the thickness and diameter of the ferromagnetic layer, and $\delta N$ the difference in dimensionless demagnetization coefficient, a.k.a., the shape anisotropy coefficient, between the perpendicular and in-plane directions. In conventional MTJs, such coefficient $\delta N$ is close to 1 (when $D >> t$), therefore, the shape anisotropy term in Eq. \ref{eq01} is usually considered a negative contribution to the PMA, which is mainly sourced by the interfacial anisotropy. However, in \textcolor{black}{PSA-STT-MRAMs}, by properly engineering the nanomagnet aspect-ratio $t/D$, the $\delta N$ can become negative, so that the shape anisotropy term, $- \delta N \frac{M_{s0}^2}{2 \mu_0} t$, provides a positive contribution to the thermal stability in synergy with the interfacial term. 

On the other hand, since the shape anisotropy strongly depends on the saturation magnetization ($\sim M_{s0}^2$), the temperature dependence of such an additional PMA source that is due to the temperature-dependent $M_s$ should not be neglected. So far, however, \textcolor{black}{due to its dominating room temperature applications}, discussions regarding the temperature-dependent behavior of \textcolor{black}{PSA-STT-MRAMs} has \textcolor{black}{not been comprehensive, with only a few pioneering reports addressing this issue \cite{lequeux_nanoscale2020,junta_apl2021}}.

\section{Temperature-dependent model}

In this work, we discuss the temperature dependence of \textcolor{black}{PSA-STT-MRAMs} by incorporating and comparing several temperature scaling models. First, we note that in Eq. \ref{eq01}, if a temperature-dependent $M_{s0}(T)$ and the Curie Temperature ($T_c$) are considered, the shape anisotropy term becomes $-\delta N \frac{M_{s0}^2 m_0^2(\tau)}{2 \mu_0} t$, where $m_0(\tau) = \frac{M_{s0}(T)}{M_{s0}(0)}$ is the reduced magnetization and $\tau = \frac{T}{T_{c}}$ is the reduced temperature. In addition, to be more complete, we also include the temperature dependence of the interfacial anisotropy, which usually follows the power scaling law, $K_i(T) = K_i(0) [\frac{M_s(T)}{M_s(0)}]^n$ according to Callen and Callen \cite{callen1,callen2}, with $M_s$ the spontaneous magnetization and $n = i(2i+1)$ corresponding to the $i^{th}-$ order anisotropy. For the $1^{st}-$ order anisotropy a scaling power $n = 3$ should be expected. However, different values of $n$ have also been reported which are related to, for example, local magnetic properties and nanofabrication related processes \cite{junta_apl2021,fu_apl2016,sato_prb2018,enobio_jjap2018,harms_srep2018,igarashi_apl2017,alzate_apl2014}. Here, we also separate the interfacial magnetization, $M_{s1}$, with the earlier bulk counterpart, $M_{s0}$, and define the reduced interfacial magnetization, $m_1(\tau) = \frac{M_{s1}(T)}{M_{s1}(0)}$, and include the temperature dependence as $K_i \propto m_1^n(\tau)$ \textcolor{black}{\cite{sun_nl2021}}. Therefore, the modified function for the thermal stability $\Delta$ can be expressed as :

\begin{equation}
\Delta =  [-\delta N \frac{M_{s0}^2 m_0^2(\tau)}{2 \mu_0} t + K_b t + K_i m_1^n(\tau)] \frac{\pi D^2}{4 k_B T}. 
\label{eq02}
\end{equation} 

\subsection{Shape Anisotropy}

Next, it is important to properly treat the temperature-dependent reduced magnetization functions, $m_0(\tau)$ and $m_1(\tau)$. To explicitly apply Eq. \ref{eq02}, the reduced magnetization should be determined empirically with appropriate fitting parameters derived from experimental measurements. However, as an evaluation of the temperature model, it is also possible to adopt the temperature scaling equation for $m_0(\tau)$ and $m_1(\tau)$ involving material dependent parameters. In treating the temperature-dependent interfacial anisotropy, the Bloch scaling law \cite{bloch_law} is usually adopted : $m_{bl}(\tau) = (1 - T/T_c)^v$, where $v$ is the scaling parameter whose value is expected to be $\sim$3/2. However, many reports in studying the temperature-dependent magnetization found that using such a single parameter cannot well reproduce the experimental results, even for $v$ values that deviate quite much from 3/2, see Fig.\ref{fig1}(b). Alternatively, a general equation for $m(\tau)$  with appropriate material-dependent parameters has been proposed by Kuz'min \textit{et al.}, which is expressed as \cite{kuzmin1,kuzmin2}: 

\begin{equation}
m(\tau) = [1 - s\tau^{3/2} - (1 - s)\tau^p]^{1/3}. 
\label{eq03}
\end{equation} 

In the above equation, the value for $p$ is 5/2 in most of the ferromagnetic materials according to the analysis from the series expansion of low-lying magnetic ex citations, and $s$ is a more material dependent parameter, with $0 < s < 5/2$, describing the functional profile of $m(\tau)$ as it varies with the reduced temperature. The theoretical derivation of $s$ depends on the intensity of the exchange interaction and the stiffness of the magnetization excitation as reveled by the Heisenberg model. In Fig. \ref{fig1}(b), we re-plot the experimental fitting curves using the Kuz'min temperature model from previous literature for a collection of materials of interests including metals, metal-alloys, Fe-(B,N) alloys, and also more recent layered van der Waals magnets \cite{yu_jac2018,kuzmin_jap2010,jian_ieee2014,dirba_acta2017,wahab_am2021,kuzmin_pla2005}. Almost all curves deviate quite much from the Bloch scaling law in which $v$ is only fitting parameter. In addition, some materials such as Nd$_5$Fe$_{17}$ exhibits quite different temperature scaling by measurements along the easy-plane (EP) and hard-axis (HA). 

These material-dependent features have been shown to be able to be properly accounted for by the parameter $s$ in the Kuz'min model. For example, the nontrivial discrepancy of the temperature-dependent magnetization in the rare-earth(RE)-FeB compounds, i.e., (RE)$_2$Fe$_{14}$B, RE = Nd, Gd, and Y, etc., has been attributed to the differences in the exchange interactions between RE-Fe as well as the stiffness around localized spin-wave branches, which are both accounted for by the parameter $s$. Last but not least, it is also found that a small modification of the Kuz'min scaling coefficient away from 1/3 can further improve the fitting to the temperature scaling of some materials \cite{wahab_am2021,kuzmin_pla2005}.

\begin{figure}[htb]
 \centering
 \includegraphics[width=3.4 in]{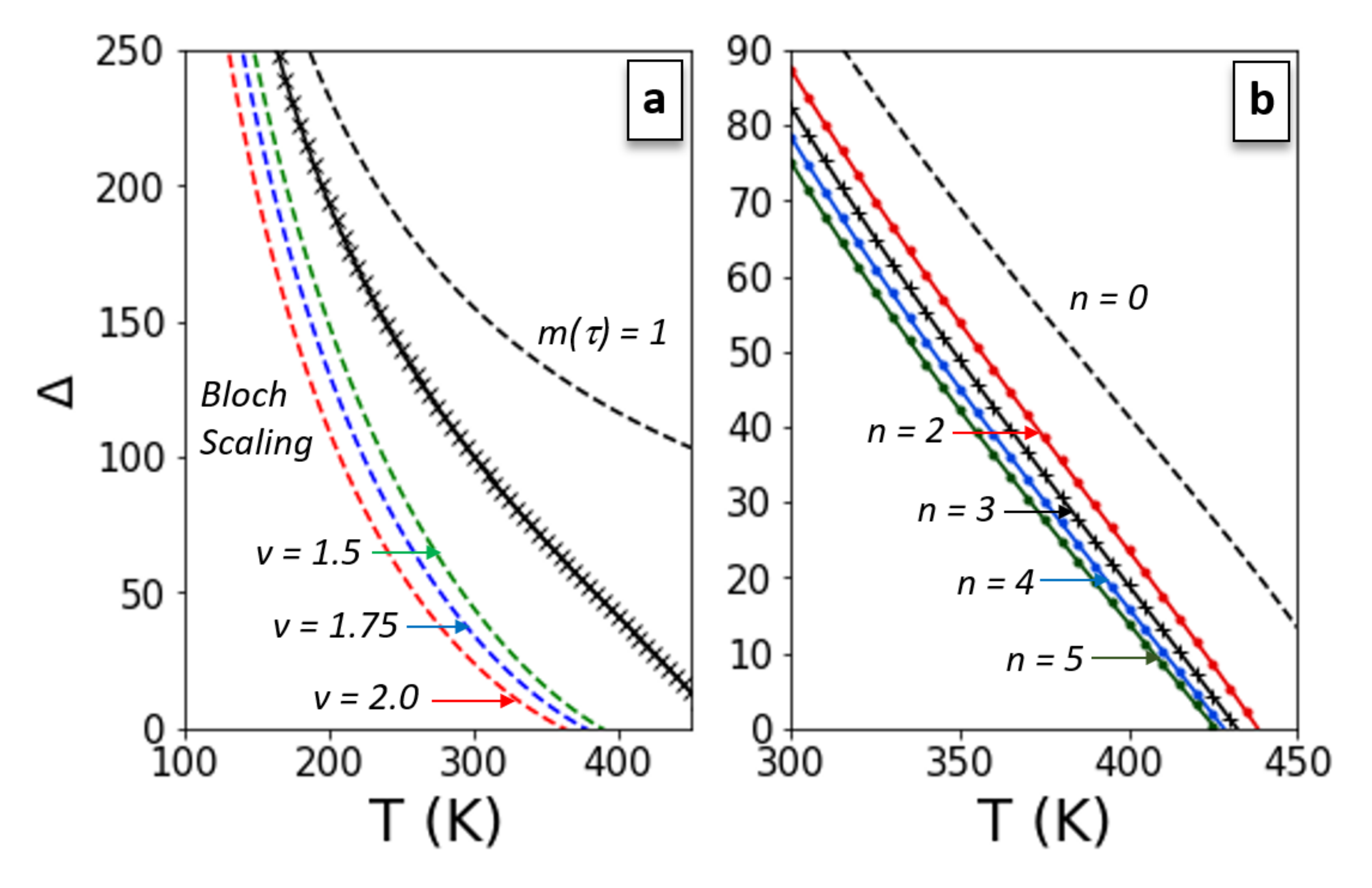}
 \caption{(a) Temperature-dependent thermal stability factor $\Delta$ for FeCoB calculated by using the Kuz'min reduced magnetization $m(\tau)$ ($\times$) and using the Bloch scaling $m_{bl}(\tau)$ (dashed line), with different $v = 1.5, 1.75, 2.0$. \textcolor{black}{Here, the temperature dependence of $K_i(T)$ is not considered. } (b) Temperature-dependent thermal stability factor $\Delta$ by using the Kuz'min scaling \textcolor{black}{[with parameters for FeCoB (0.65,2.5)]} for both $m_0(\tau)$ and $m_1(\tau)$, at different interfacial anisotropy coefficient $n = 0, 2, 3, 4, 5$.  }
 \label{fig2}
\end{figure}

In order to demonstrate the temperature-dependent effects, we adopt the Kuz'min scaling law and use the parameter for FeCoB ($s = 0.65, p = 5/2$), and realistic values of $T_c = 480$ K, $M_{s0} = 1.52$ T, $K_b = -1.1 \times 10^5$ J m$^{-3}$, and $K_i = 2.2 \times 10^{-3}$ J m$^{-2}$, as adopted from previous literature. The $T_c$ value we use is on the lower end, to also take into account the finite size effect \cite{zhang_prl2001} which is relevant to the nanopillars with dimensions $\sim$ tens of nm$^2$ or even smaller. It is also noted that these values should be only viewed as one set of example values, as the parameters, including the $T_c$, can be largely different upon materials and devices engineering. \textcolor{black}{We present in} Fig. \ref{fig2}(a) the temperature-dependent thermal stability factor calculated by using the different temperature scaling laws \textcolor{black}{and without considering the temperature dependence of $K_i(T)$}.  As an example, the dimensions of the nanopillar free-layer is set to $t = 30$ nm and $D = 10 $ nm (aspect-ratio: $t/D = 3$). More details regarding the nanomagnet's dimension dependence will be discussed later. 

First, we note that if the temperature-dependent $M_{s}$ is not considered \textcolor{black}{ [$m(\tau) = 1$]}, a significant overestimation of $\Delta$ can be expected compared to that using the Kuz'min scaling. At around room temperature ($\sim$ 300 K), the difference can be as large as $\delta \Delta \approx $ 50. Such a discrepancy further increases as temperature increases, which can be close to $\delta \Delta \approx $ 100 near the Curie temperature of FeCoB. This result shows the importance of considering the temperature-dependent magnetization in analyzing \textcolor{black}{PSA-STT-MRAMs}, especially for higher temperature related applications ($\sim$ 150 $^\circ$C). On the other hand, the Bloch scaling seems to always underestimate $\Delta$, regardless of the chosen $v$ (example curves are shown for $v$ = 1.5, 1.75, and 2). The comparison between the Bloch scaling and the Kuz'min model indicates the important role of the material-dependent parameter $s$ in determining the energy barrier between the two states.  

\begin{figure*}[htb]
 \centering
 \includegraphics[width=5.8 in]{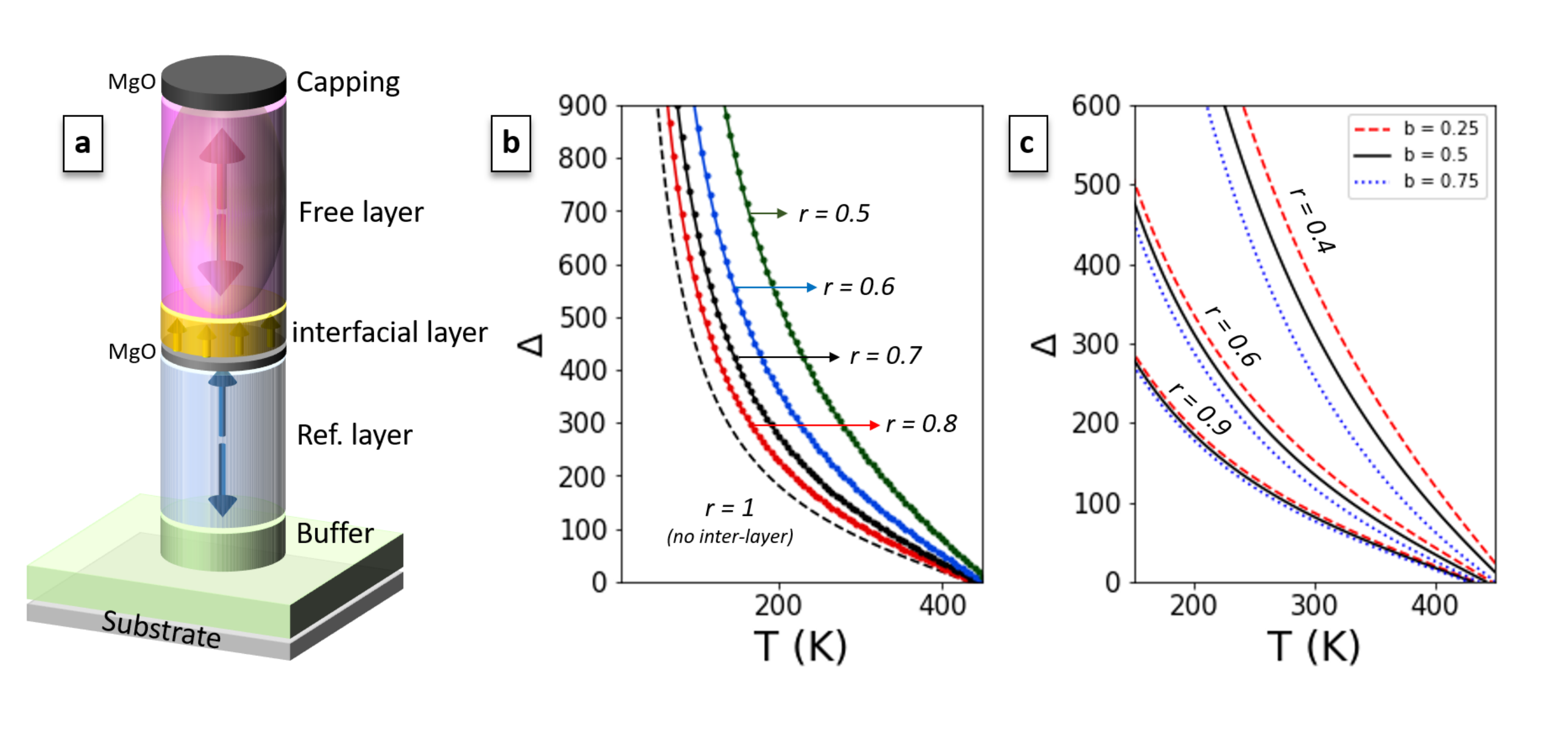}
 \caption{(a) Schematic illustration of the \textcolor{black}{PSA-STT-MRAMs} after considering an interfacial layer with an attenuated saturation magnetization $M_{s1} = r \times M_{s0}$, for the temperature scaling of the interfacial anisotropy. (b) Temperature-dependent thermal stability factor $\Delta$ at different attenuation coefficient $r$, where $r = 1$ corresponds to no inter-layer magnetization. (c) Temperature-dependent thermal stability factor $\Delta$ at different thermal coefficients $b = $ 0.25, 0.5, and 0.75, for three $r$ values ($r = 0.4, 0.6, 0.9$). }
 \label{fig3}
\end{figure*}

\subsection{Interfacial Anisotropy}

Figure \ref{fig2}(b) compares the different temperature-dependent $\Delta$ caused by the interfacial anisotropy $K_i(T)$. Here, the Kuz'min model for the reduced magnetization is used for all the plotted curves (same shape anisotropy), however, different values of the interfacial anisotropy scaling coefficient ($n$) are compared, in which $n = 0$ (dashed line) indicates a temperature-independent $K_i$. Other example curves for $n = 2,3,4,5$ are shown, where $n = 3$ is expected from theory for the $1^{st}-$ order anisotropy.  Experimental studies have shown that the $m^3(\tau)$ scaling is generally successful in predicting the temperature dependence of interface-driven p-MTJs such as FeCoB/MgO and Fe/MgO. The temperature-dependent $K_i(T)$ only becomes significant to the $\Delta$ at around room temperatures and up to $T_c$. In general, an overestimation on $\Delta$ (in the range of $\delta \Delta \sim 10 - 20$) can be expected if one does not consider any temperature-dependent $K_i$ ($n = 0$), see Fig. \ref{fig2}(b). In addition, the different choices of $n$ only accounts for a small modification to the $\Delta$, around $\delta \Delta < $ 10. However, \textcolor{black}{since interfaces are generally rather sensitive to extrinsic thin-film parameters, the actual temperature dependence of interfacial MTJs is usually underestimated in a theoretical modeling. Furthermore, abrupt temperature-driven property changes may also arise \cite{dieny_jphysd2019}, which may be attributed to multiple magnetic phases and/or exchange bias at the interface. On the other hand, the temperature-dependent shape anisotropy often reflects a more intrinsic dependence with the material properties, i.e. saturation magnetization, which makes the temperature-dependent properties of PSA-MTJs more reliable in performance and predicable with theoretical modeling. \cite{dieny_jphysd2019,lequeux_nanoscale2020,junta_apl2021} }

Another critical issue that would affect the temperature scaling related to the $K_i$ contribution is whether a separated interfacial magnetization, $M_{s1}$ should be considered, which is different from the bulk $M_{s0}$, as illustrated in Fig. \ref{fig3}(a). This treatment makes general sense, since many layer-resolved studies have indicated the existence of an attenuated interface/surface magnetization value ($M_{s1}$) or even a dead-layer,  and the anisotropy is of interfacial origin that should then naturally scale only with the $M_{s1}$. As an example, such a consideration has recently found a great relevance in the temperature scaling of Fe/MgO interfacial p-MTJs \cite{ibrahim_arxiv2020}. In \textcolor{black}{PSA}-MTJs, the separation of $M_{s0}$ and $M_{s1}$ seems more critical as both the shape ($ \propto m^2_{0}$) and interfacial ($\propto m^n_{1}$) anisotropies contribute to the thermal stability factor.

Here, we take into account an interfacial magnetization that is attenuated by a factor of $r$ with respect to the bulk value, i.e. $r = \frac{M_{s1}}{M_{s0}}$, to properly account for the temperature scaling of $K_i$. For simplicity, we adopt the Kuz'min scaling for both $m_{0}(\tau)$ and $m_{1}(\tau)$, however, it is noted that the most explicit function should be determined empirically with fitting parameters derived from experimental measurements, as the Kuz'min scaling parameter may also be different for the bulk and interfacial components.

\begin{figure*}[htb]
 \centering
 \includegraphics[width=5.5 in]{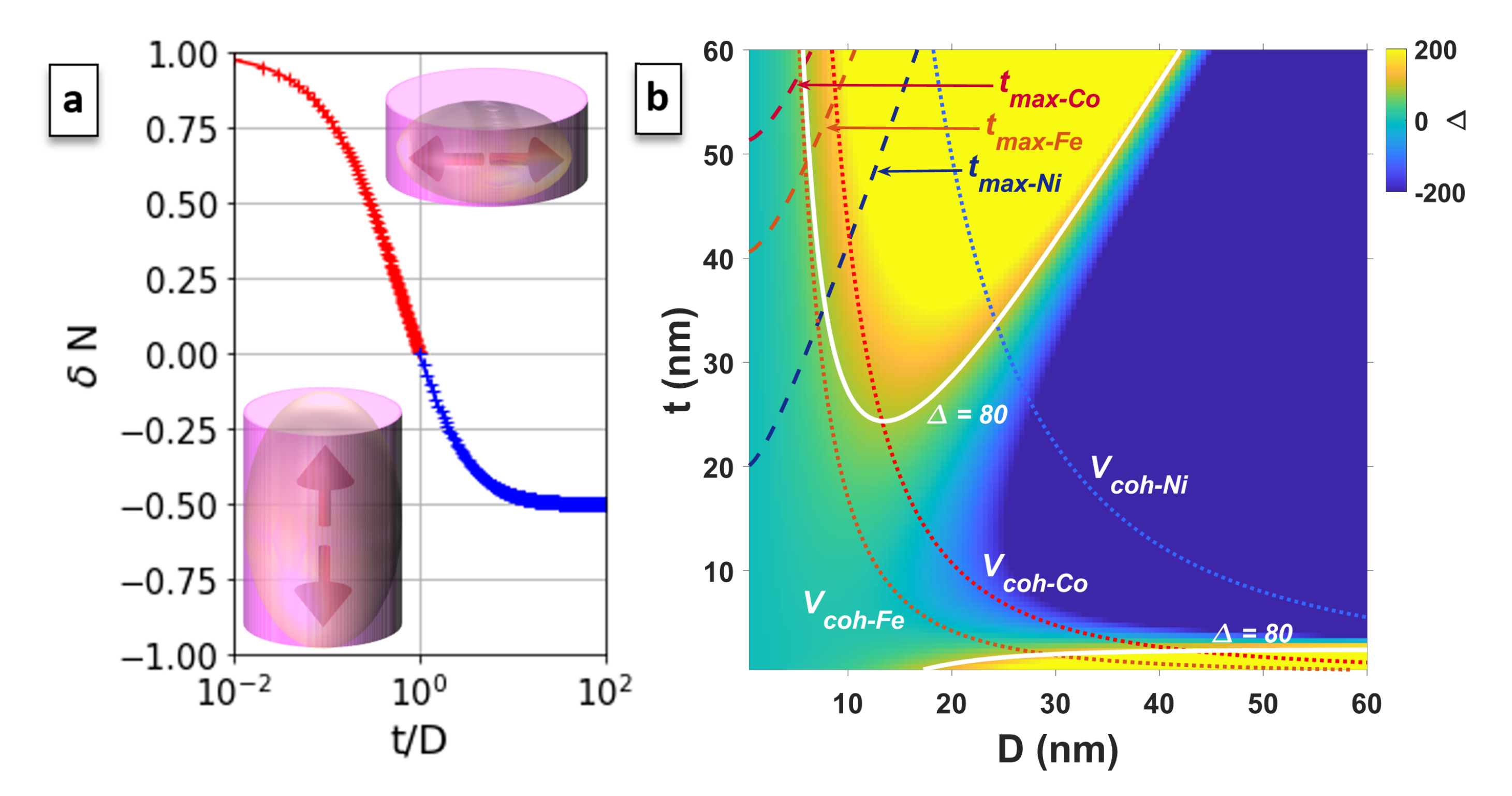}
 \caption{(a) Calculated shape anisotropy coefficient $\delta N = N_z - N_x$ as a function of the nanomagnet dimension ($t, D$). For $t/D <1$, the oblate spheroid case is used and for $t \ge D$, the prolate spheroid case is used. (b) Mapping of the thermal stability factor $\Delta$ as a function of the nanomagnet dimension ($t, D$) at $T = 300$ K. Scale-bar: $[-200, 200]$. The calculated $t$ versus $D$ corresponding to the coherence volume \textcolor{black}{(dotted) and the maximum macrospin thickness (dashed)} of Ni, Co, and Fe are also plotted. }
 \label{fig4}
\end{figure*}

Figure \ref{fig3}(b) shows the effect of such an attenuating factor $r$ on the temperature scaling of $\Delta$, where $r = 1$ indicates $M_{s0} = M_{s1}$, and $r = 0$ represents an interfacial dead-layer. The separation of interfacial and bulk magnetization seems quite influential. Not considering the interfacial magnetization effect leads to an overall underestimation of $\Delta$. In general, this effect becomes rather significant at lower temperatures. Nevertheless, at around room temperature, a non-trivial modification of $\Delta$ about $50 - 100$ is still predicated (for a practical range of $r$, $\sim 0.6 - 0.7$).   

The inclusion of the surface magnetization reduction also brings up a necessary discussion of the thermal expansion effect to the temperature dependent scaling. Due to the sputtered hetero-junction formed between the thin FeCoB layer and the MgO barrier, the interfacial PMA can become quite sensitive to lattice strain \cite{li_apl2015}. In fact, strain engineering has been a key topic among other practical solutions for bench-marking MTJ performance \cite{zhao_apl2016,loong_srep2014}. Such an interfacial strain effect would become more significant in thinner MTJ devices. Resultantly, the temperature dependent thermal expansion effect should non-trivially factor in to the temperature dependent interfacial anisotropy, especially when considering a magnetization reduction at the interface ($M_{s0} \ne M_{s1}$). Usually, the temperature dependent thermal expansion effect follows a scaling behavior of $(1 - b \frac{T}{T_c})$, where $b$ is the thermal expansion coefficient \cite{carr_pr1958, lee_aipadv2017,lequeux_nanoscale2020}. Such a correction factor should be superimposed to the existing Callen-Callen law for interfacial anisotropy, therefore, the temperature dependent interfacial component becomes: $K_i(T) = K_i (1 - b \frac{T}{T_c}) m^n_1(\tau)$. 

We substitute this modified $K_i(T)$ contribution to our analysis in Eq. \ref{eq02}.  Figure \ref{fig3} (c) shows the effect of different thermal expansion coefficients $b$ at three different $M_{s1}$ attenuation ratios ($r =$ 0.9, 0.6, and 0.4). First, we note that the thermal expansion effect to the calculated $\Delta$ is most dramatic at intermediate to high (approaching $T_c$) temperature ranges, e.g., $T =$ 200 $-$ 400 K. At lower temperatures, the shape anisotropy become more dominant ($\sim M^2_{s0}$), rendering the surface magnetization effects nearly negligible. Second, the effective modification to $\Delta$ increases as the attenuation ratio decreases. When $r$ is large, the thermal expansion effect is nearly negligible, however, for smaller $r$ values such as $r = 0.4$, a discrepancy of $\delta \Delta > 50$ can be obtained via changing $b$ from 0.25 to 0.75, at $\sim 300$ K.

\textcolor{black}{Last but not least, we note that one other contribution which have not been accounted for is the temperature-dependent magnetocrystalline anisotropy. For soft ferromagnets with significant $\delta N$, like in the case of \textcolor{black}{PSA-STT-MRAMs}, the magnetocrystalline anisotropy usually plays a less critical role and therefore is generally neglected. However, for materials with large magnetocrystalline anisotropy, the temperature dependence should further accounts for the effect of $K_b(T)$, and a nontrivial extra correction may be then expected besides the shape and interfacial anisotropy effect considered herein.}

\section{Switching mechanism, coercivity, and Blocking temperature}

Next, we shift our focus to the temperature-dependent coercivity. Unlike the thermal stability factor, the coercivity is a more extrinsic parameter that depends not only the magnetic anisotropy, but also the dimension and shape of the nanomagnet, as well as the magnetization reversal mechanisms, such as coherent rotation or domain wall nucleation and propagation. In the context of SW particles, the volume size for a coherent rotation satisfies $V < V_{coh} \sim (L_{coh})^3$, in which $L_{coh}$ is the coherence length. For Ni, Co, and Fe, the coherence lengths are 25, 15, and 11 nm, respectively \cite{sellmyer_jpcm2001, skomski_jpcm2003, he_prb2007}. As a result, to properly model the coercivity, the volume size of the free-layer relative to the coherence volume needs to be taken into account. To this purposes, we define a reduced volume, $V_{red} = V/V_{coh}$.

\subsection{Free-layer Dimensions}

First, for a given volume size, $V = \frac{\pi D^2}{4}t$, the nanomagnet dimension ($t, D$) directly impacts the shape coefficient $\delta N$. For a nanopillar free-layer, the demagnetization coefficients $N_x = N_y$ and $N_x + N_y + N_z = 1$, therefore $\delta N = N_z - N_x$, where $z$ is along the film normal direction. In Fig. \ref{fig4}(a), we plot the shape coefficient $\delta N$ following the theoretical equations for two cases: the oblate spheroid ($t<D$), and prolate spheroid ($t \ge D$), which have been used in describing the \textcolor{black}{PSA-STT-MRAMs} design window in many earlier reports \cite{watanabe_ncomm2018,dieny_jphysd2019,jinnai_apl2020,jinnai_iedm2020}. 

Figure \ref{fig4}(b) shows the thermal stability factor $\Delta$ as a function of $t$ and $D$ at $T = 300$ K, using \textcolor{black}{$T_c = 480$ K, $M_{s0} = 1.52$ T, $K_b = -1.1 \times 10^5$ J m$^{-3}$, and $K_i = 5.0 \times 10^{-3}$ J m$^{-2}$}. It features two high-$\Delta$ regions: (i) one at the bottom-right corner that corresponds to the conventional interfacial anisotropy MTJ. Here, high enough $\Delta$ (usually $\Delta > 80$) may not be obtained for smaller $D$ values such as $D < 20$ nm. (ii) A much larger one near the top that corresponds to the \textcolor{black}{PSA}-MTJ. Here, high enough $\Delta$ can be realized even for $D$ values $\sim 10$ nm or less thanks to the shape anisotropy contribution. Separating the above two high-$\Delta$ regions is a dark region, which represents the case for in-plane shape anisotropy.

In Fig. \ref{fig4}(b), we also overlay curves representing the coherence volume, $V_{coh}$, using the values for Ni, Co, and Fe, respectively. Under a certain $V_{coh}$ curve the magnetization reversal satisfies the coherent rotation criteria, and above it, domain wall processes or a mixed reversal mechanism may be relevant. From our calculation, it is seen that the \textcolor{black}{PSA}-MTJ region from thermal stability point of view has an appreciable overlap with the coherent rotation region for $V_{coh-Ni}$, however, only a small overlap for $V_{coh-Fe}$ and $V_{coh-Co}$. This suggests that the domain-wall-driven magnetization reversal may still be largely relevant in (Fe,Co)B \textcolor{black}{PSA-STT-MRAMs}. However, such a scenario could easily change via proper engineering of material parameters including the anisotropy energies and the coherence length. Besides, in order to maintain a certain volume size, one needs to increase the $t$ significantly if the $D$ has to decrease due to, for example, ultrahigh density considerations. 

\textcolor{black}{In addition, another constrain comes directly from the aspect ratio ($t/D$) affecting the thermal stability factor due to the limitation of the macrospin approach, i.e., there is an upper limit of thermal stability when we increase the $t/D$, since the magnetization, beyond such a threshold $t/D$, would favor a reversal via domain wall nucleation and propagation \cite{dieny_nanoscale2018}. Therefore, we estimate the maximum macrospin thickness, $t_{max}$, as a function of the MTJ diameter using the method presented in Perrissin \textit{et al.}\cite{dieny_nanoscale2018} and plot three curves for Ni, Fe, and Co in Fig.\ref{fig4}(b), respectively. We can see from Fig.\ref{fig4}(b) that we need to consider both the coherence volume and the maximum macrospin thickness effects whenever we attempt to utilize MTJs with large PSA ($\Delta > 80$), especially for Co- and Fe-based MTJs. Last but not the least, to mitigate the maximum macrospin thickness issue, we may also adopt the method of inserting a thin non-magnetic layer in the free layer \cite{jinnai_iedm2020}.}

\subsection{Temperature-dependent Coercivity}

To model the temperature-dependent coercivity, $H_c(T)$, we write the temperature-dependent anisotropy, $K(\tau) = \Delta \times k_B T_c \tau /V$, and therefore, $H_c(T) = \frac{\mu_0 g K(\tau)}{M_{s0} m_0(\tau)} \times \Big\{ 1 - \left[ \frac{\mathrm{In}[t/t_0]}{\Delta(\tau)} \right]^{1/\alpha} \Big\}$, where $\mathrm{In}[t/t_0]$ is a factor related to the time necessary for jumping over the energy barrier and is usually estimated as about 25, $g$ and $\alpha$ are fitting parameters related to the anisotropy axis and the size of the nanomagnet. We use $g = 2, \alpha = 2$ to approach the magnetization reversal in the context of SW particle with a coherent rotation mode. 

\begin{figure}[htb]
 \centering
 \includegraphics[width=3.0 in]{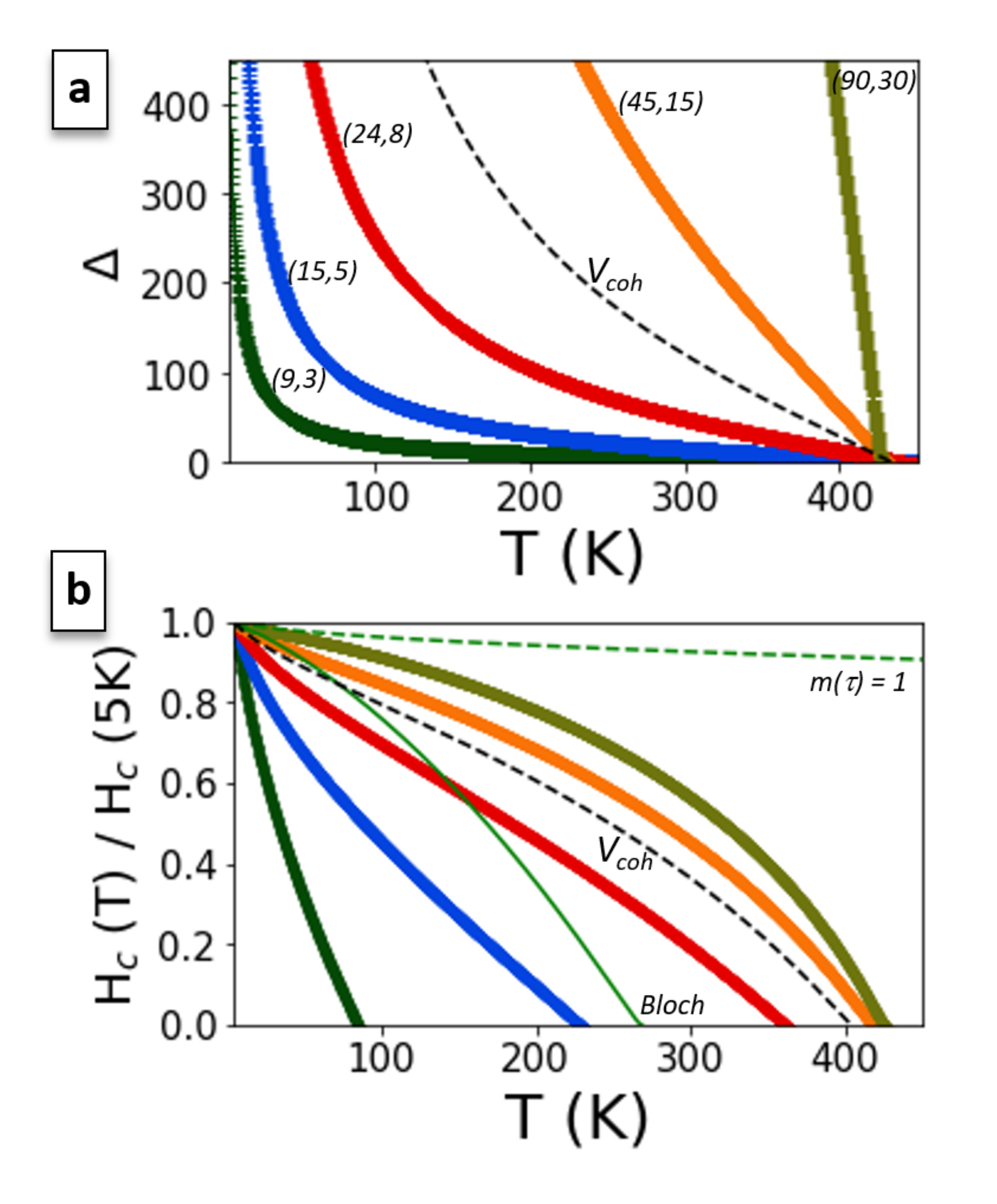}
 \caption{(a) Temperature-dependent thermal stability factor, $\Delta$ calculated for selective nanopillar dimensions ($t, D$), $D = 3, 5, 8, 15, 30$ nm, with a fixed aspect-ratio ($t/D = 3$). (b) The corresponding calculated temperature-dependent coercivity, $H_c(T)$. The dashed lines in (a) and (b) represent the results obtained using the coherence volume of Co. For $D = 30$ nm, the calculated $H_c(T)$ for $m(\tau) = 1$ and for $m(\tau)$ using the Bloch function are also plotted as references. }
 \label{fig5}
\end{figure}

In Fig. \ref{fig5}(a), we plot the temperature-dependent thermal stability factor for selective nanopillar dimensions ($D = 3, 5, 8, 15, 30$ nm) with a fixed aspect-ratio ($t/D = 3$). A curve representing the Co coherence volume is also shown (dashed line) as a reference. Figure \ref{fig5}(b) shows the corresponding temperature-dependent coercivity referenced to a low temperature value $H_c(5 \mathrm{K})$. The temperature-dependent $\Delta$ changes rapidly with the increase of the volume size and is basically irrelevant to the critical coherence volume. For the coercivity, the temperature profile $H_c(T)$ also exhibits a strong dependence on the nanomagnets' dimension in the coherent rotation region ($V < V_{coh}$), especially for the blocking temperature ($T_B$), at which the $H_c(T)$ crosses zero. However, the $H_c(T)$ profile, as well as the $T_B$ value, does not change much with further increasing the dimension, once the volume size exceeds the coherence volume. In addition, the Bloch function of $m(\tau)$ (solid line) also cannot well reproduce the $H_c(T)$ profile, and it tends to underestimate the blocking temperature, especially at regions beyond the coherence volume.  

\begin{figure}[htb]
 \centering
 \includegraphics[width=3.3 in]{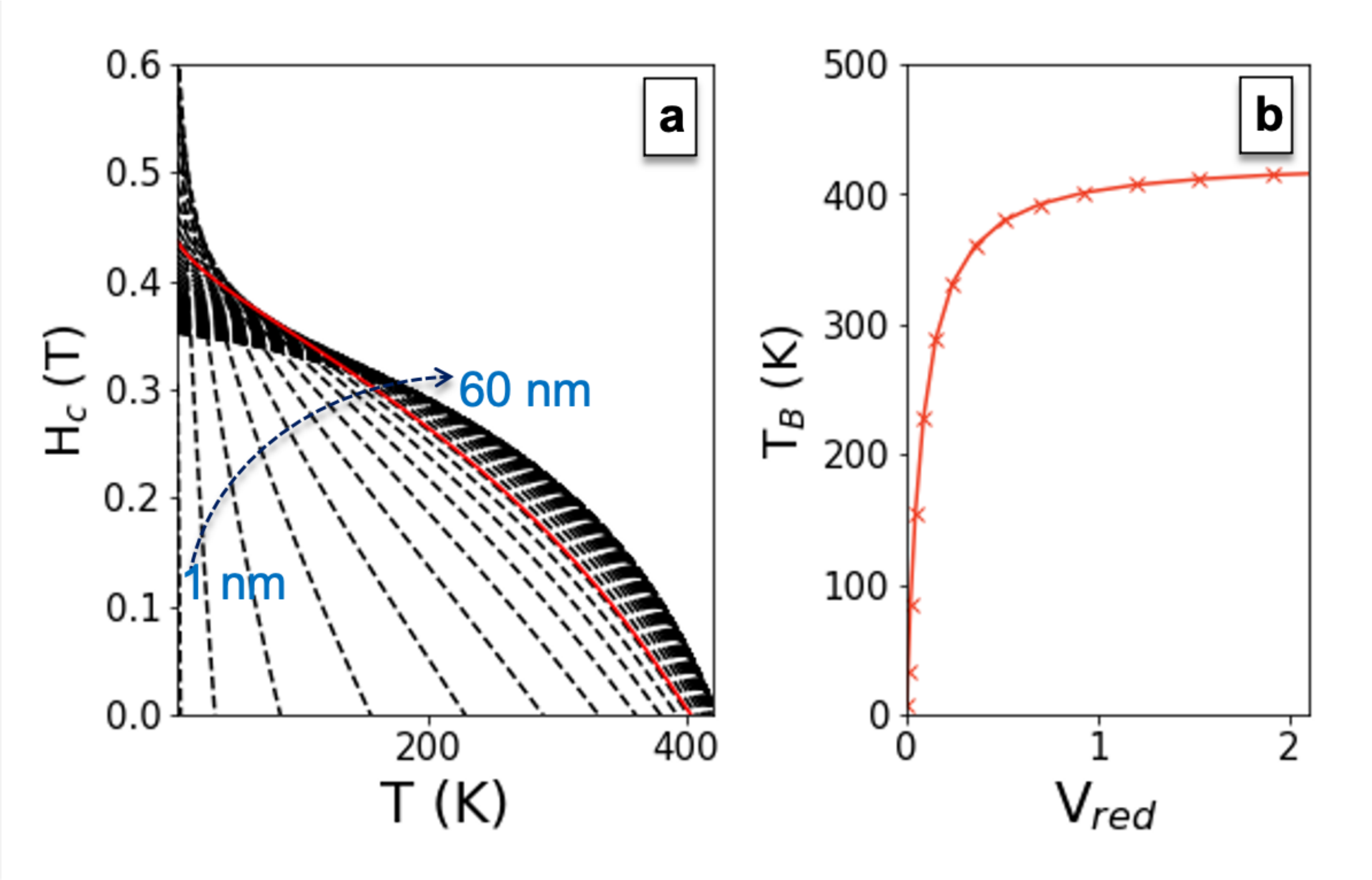}
 \caption{(a) (Dashed line) Temperature-dependent coercivity $H_c(T)$ at a range of nanopillar dimension $D$ increasing from 1 to 60 nm with a step size of 1 nm. The aspect-ratio is kept as $t/D = 3$. (Solid line) $H_c(T)$ reference calculated for a nanopillar dimension corresponding to the coherence volume of Co. (b) The numerically-solved blocking temperature, $T_B$, from (a), versus the reduced volume $V_{red} = V/V_{coh}$, at an fixed aspect-ratio $t/D = 3$. }
 \label{fig6}
\end{figure}

\subsection{Blocking temperature}

Recently, it has been found interesting and technological relevant to study MTJs that operate in the superparamagnetic regime, a.k.a. stochastic MTJs, near and above the MTJs' blocking temperature. The fluctuation of magnetization in stochastic MTJs can be used for high-dimensional optimization or sampling problems in probabilistic computing \cite{sun_nl2021,fukami_prl2021}.     
Usually, the blocking temperature for SW particles with small volume size $V$ and small shape coefficient $\delta N$ is proportional to the potential barrier that only weakly depends on the temperature effect of magnetization. However, for larger particles with larger shape anisotropy, as in the present case, the correction factor attributed to the temperature variation becomes significant. This effect can be already observed in Fig. \ref{fig5}(b) for the different nanomagnet dimensions. This is because the correction on $H_c(T)$ attributed to the temperature-dependent shape anisotropy is realized by the reduced magnetization $m(\tau)$, which directly results in the correction factor $m(\tau_B)$, where $\tau_B$ is the reduced blocking temperature. At lower temperatures, $m(\tau) \approx 1$, a constant energy barrier is a good approximation to estimate the blocking temperature. On the other hand, if $m(\tau)$ varies significantly from $m(0) = 1$, such as near the $T_c$, then the determination of $T_B$ cannot neglect the correction effect arising from the $m(\tau)$.

To further evaluate the dimension effect on the blocking temperature, we calculated the $H_c(T)$ for a range of $D$ values, as shown in Fig. \ref{fig6}(a), while keeping the aspect-ratio the same ($t/D$ = 3) to ensure a same $\delta N$. A curve representing the coherence volume is also plotted as a reference (solid-line) to indicate the coherent rotation region. The blocking temperature, as derived from the coercivity function, is described as: $T_B = E(\tau_B) / [k_B \mathrm{In}(t/t_0)]$. The dependence of $T_B$ on the reduced volume, $V_{red}$, can be then solved numerically. the result is plotted in Fig. \ref{fig6}(b). In the limit of small volume size, e.g. $V_{red} \sim 0.1 - 0.3$, $T_B$ is small and varies almost linearly with $V_{red}$, since $m(\tau_B)$ approaches 1 as $\tau_B$ approaches 0. As the nanopillar volume grows beyond $V_{red} \sim 0.3$, $\tau_B$ begins to deviate from the linear behavior with respect to $V_{red}$, since $m(\tau_B)$ decreases nonlinearly towards 0 and the correction to $m(\tau_B)$ becomes more significant. Such a nonlinear behavior also depends on the nanomagnet dimension (for a given $V$), and is more pronounced for higher aspect-ratio nanomagnet.  

\begin{figure*}[htb]
 \centering
 \includegraphics[width=5.7 in]{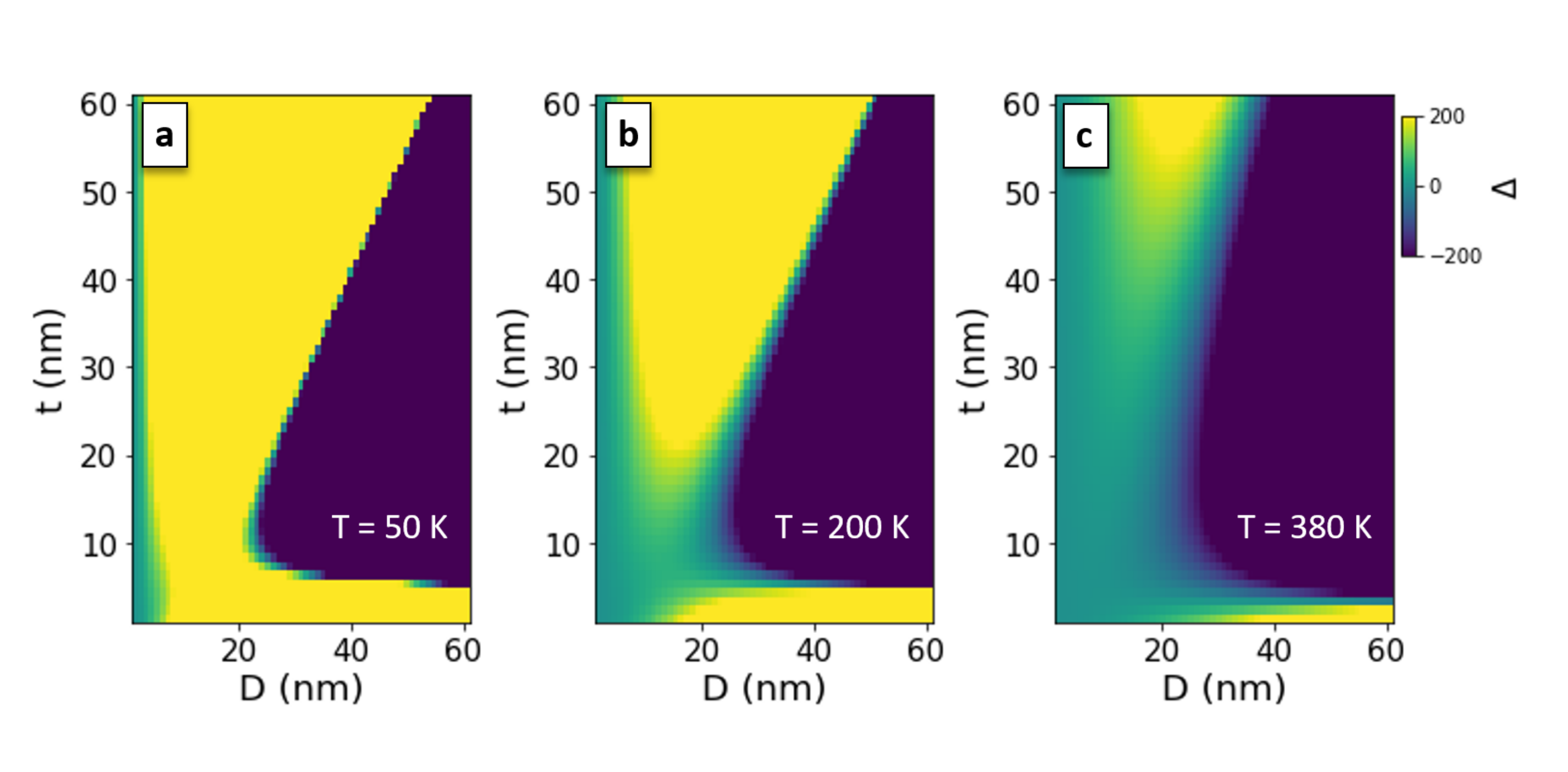}
 \caption{The thermal stability factor ($\Delta$) as a function of the nanomagnet dimension ($t,D$) at selective temperatures: (a) 50 K, (b) 200 K, and (c) 380 K. Scale bar: [$-200,200$]. }
 \label{fig7}
\end{figure*}

\textcolor{black}{\subsection{Current-driven Switching}}

Considering the potential applications of \textcolor{black}{PSA-STT-MRAMs} at both lower and higher temperatures, we further discuss and compare the thermal stability factor map $\Delta (t,D)$ at selective temperatures, $T = 50, 200, 380$ K, in Fig. \ref{fig7}. At $\sim 50$ K and below (down to the cryogenic temperature), the shape- and interfacial-induced stability merge together below $D \sim $ 20 nm. In addition, it seems practical to secure a good thermal stability even for a nanopillar dimension below 10 nm. The overlap of the two regions indicate a synergistic effect of the shape and interfacial anisotropies. As temperature increases, both regions shrink at about the same rate. However, it is still possible to identify a reasonably large design window for different $(t,D)$ combinations at a wide range of temperature from 100 - 300 K. At higher temperatures, the acceptable range for $D$ becomes narrower (centered around 20 nm) and higher aspect-ratio of the nanomagnet is also desirable.

\begin{figure}[htb]
 \centering
 \includegraphics[width=3.4 in]{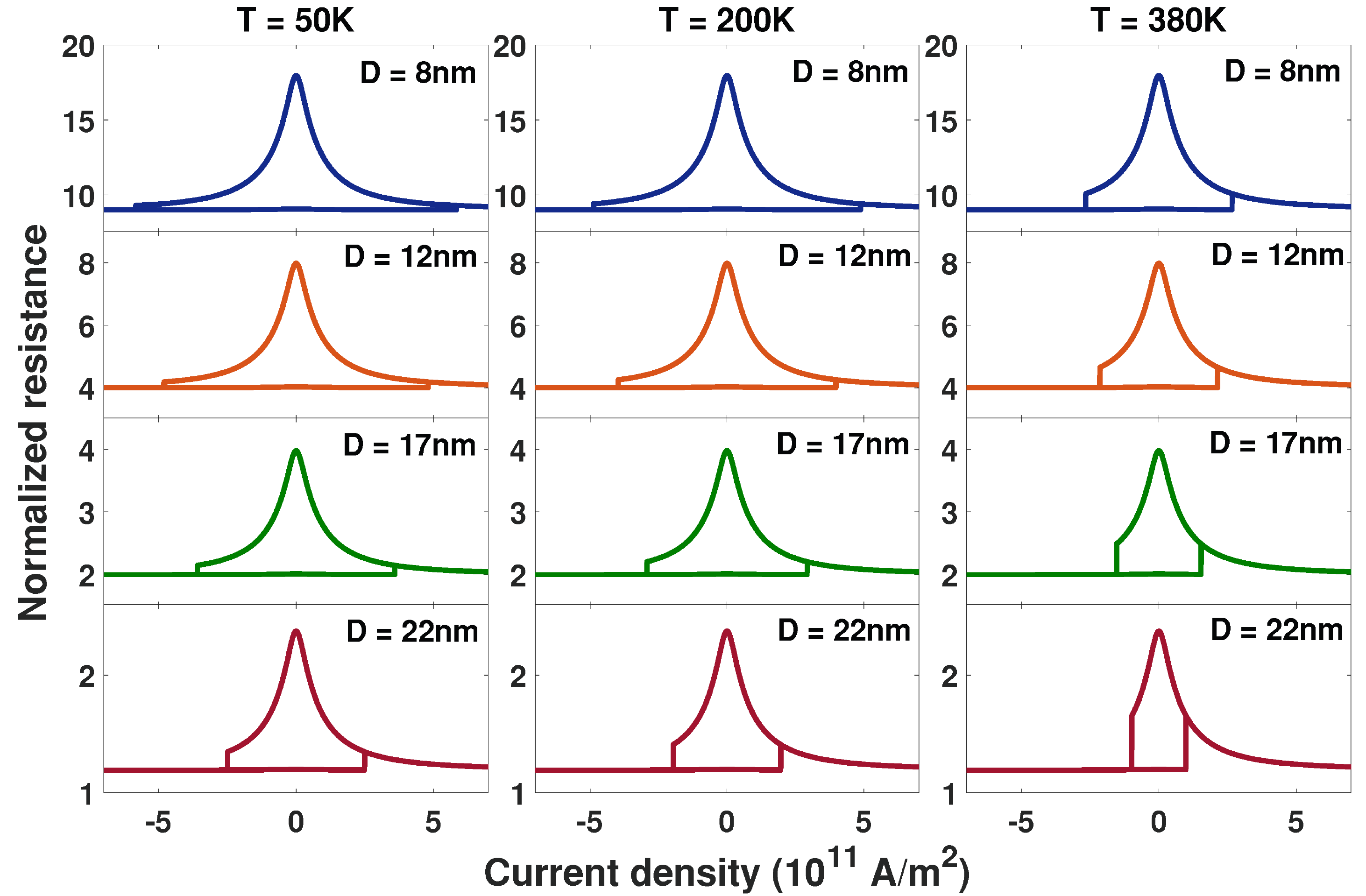}
 \caption{The STT switching loops as a function of the MTJ diameter ($D$) at selective temperatures: (a) 50 K, (b) 200 K, and (c) 380 K. The MTJ thickness is set to 30 nm. }
 \label{fig8}
\end{figure}

We also evaluate the critical current density ($J_{c0}$) as a function of the MTJs' dimension and the temperature in a current switching scenario via the  spin-transfer torque (STT). The critical current density, $J_{c0}=8\alpha e \gamma  \Delta  k_B T/(\mu_B g_{\text{STT}}) $, where $\gamma$ is the gyromagnetic ratio, $e$ is the electron charge, $\alpha$ is the damping constant, $\mu_B$ is the Bohr magneton and $g_{\text{STT}}$ is the STT efficiency \cite{wszhao_ieee2018}. The $g_{\text{STT}}$ is given by $p/2/(1+p^2\text{cos} \theta)$ in the single reference layer MTJ and $p/(1-p^4\text{cos}^2 \theta)$ in the dual reference layer MTJ, where $p$ is the spin polarization factor and $\theta$ is the initial angle between the magnetizations of the free layer and reference layer \cite{diao_apl2007}. We use dual reference layer structure to perform our calculations. To obtain the STT switching curves, we normalize the resistance using the parallel resistance of the \textcolor{black}{PSA-STT-MRAM} structure, and therefore, only need to consider the voltage-dependent tunnel magnetoresistance ratio (TMR). The TMR is given by $\text{TMR}(0)/[1+(V_{sw}/V_h)^2]$, where $V_{sw}$ is the applied (bias) voltage across the MTJ and $V_h$ is the bias voltage where TMR($V_h$) is half of TMR(0). As an example, we show the MTJ-diameter-dependent STT switching loops for different temperatures in Fig. \ref{fig8}.

\begin{figure*}[htb]
 \centering
 \includegraphics[width=5.5 in]{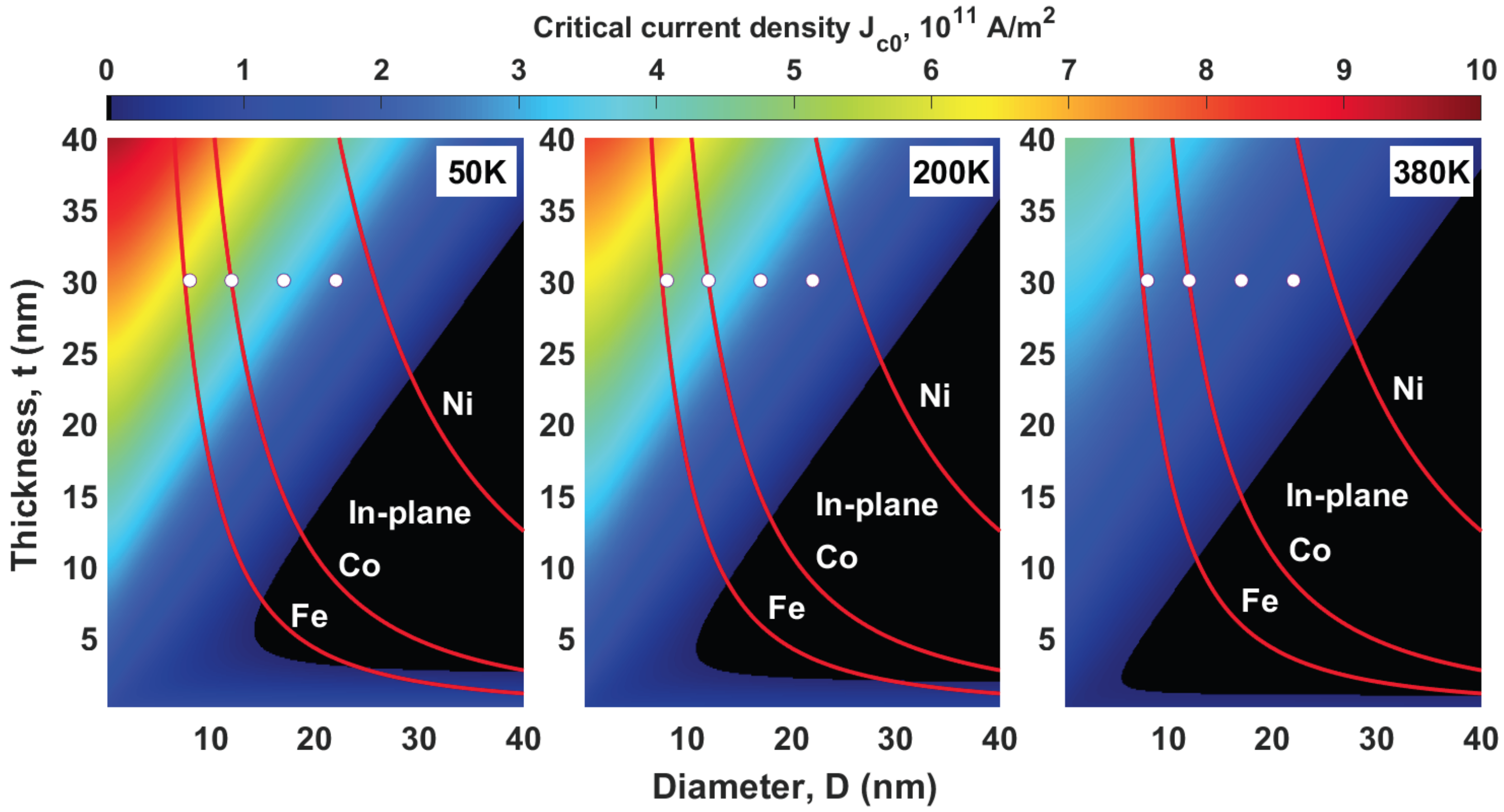}
 \caption{The critical current density as a function of the nanomagnet dimension ($t,D$) at selective temperatures. The MTJ thickness is 30 nm. The calculated $t$ versus $D$ corresponding to the coherence volume of Ni, Co, and Fe are also plotted. The four dots correspond to the four MTJ dimensions discussed in Fig. \ref{fig8}.}
 \label{fig9}
\end{figure*}

The STT switching simulations capture the essential physics since the simulated curves are qualitatively similar to the experimental observations \cite{watanabe_ncomm2018}. We further plot the critical current density map $\Delta (t,D)$ at selective temperatures, $T = 50, 200, 380$ K, in Fig. \ref{fig9}. The four different sizes presented in Fig. \ref{fig8} are also indicated on the map. At a fixed size, or a fixed aspect-ratio, the switching current density of the MTJ nanomagnet strongly depends on the temperature. For example, for a nanomagnet dimension of (30, 8), the switching current can be a few times different as one goes from cryogenic temperature to above room temperature.  

Similar to our analysis in the thermal stability factor map, we again overlay the coherence volume curves, using the values for Ni, Co, and Fe, respectively. Under the curve, the magnetization reversal  satisfies  the  coherent  rotation  criteria without necessarily triggering the domain wall processes. From our calculation, it is seen that the coherent current-switching model is still limited to the smaller nanomagnet dimensions in the \textcolor{black}{PSA-STT-MRAM} window, which  suggests that the domain-wall-driven magnetization reversal may be largely relevant in the current-driven switching scenario. \textcolor{black}{Further, we notice from a recent report that the switching could be driven by domain nucleation and domain wall motion even when the magnet size is small, but as long as the $t/D$ is larger than one \cite{Cacoilo_arxiv2021}. The corresponding temperature dependent behaviors with respect to these domain wall processes, though beyond the scope of the current work, may deserve more future investigations.}

\section{Summary and Discussions}

In summary, we provided a model to analyze the temperature dependence of shape-anisotropy magnetic tunnel junctions. The thermal stability factor, described in Eq.\ref{eq01} is modified to include the temperature dependent shape and interfacial anisotropy, Eq.\ref{eq02}, which is particular important for nanopillar-shaped free-layer of ferromagnetic materials. The corrections on field-switching coercivity, $H_c(T)$, blocking temperature, $T_B$, and current-driven switching using STTs are subsequently derived and presented, which attributes to the effect of temperature dependent spontaneous magnetization $m(\tau)$. 

\textcolor{black}{Finally, we note that another important effect which have not been accounted for in this work is} a quantitative analysis regarding nanomagnet's finite size effect. The finite size effect of nanostructures is mainly related to the geometrical confinement to the correlation length, causing a reduction in the ordering temperature $T_c$. In addition, this effect will be further modified by the free surface effect as the nanomagnet size approaches the ultrafine limit, smaller than the effective range of spin-spin interaction \cite{zhang_prl2001}. Last but not the least, we point out that in many electric-current-induced switching experiment, additional temperature-dependent parameters need to be further enclosed, such as the spin-transfer torque efficiency, magnetic damping, resistance-area product, and so on. However, these factors as well as their temperature behaviors may be worth of a separated study.

\textcolor{black}{ \section*{Acknowledgment}
The work at Oakland University was supported by U.S. National Science Foundation under Grants No. ECCS-1933301 and ECCS-1941426. The work at University of Arizona was supported by U.S. National Science Foundation under Grants No. DMR-1905783. The work at HKUST was supported by UROP program and HKUST-Kaisa Joint Research Institute grant (OKT21EG08).}

\textcolor{black}{ \section*{DATA AVAILABILITY}
The data that support the findings of this study are available from the corresponding author upon reasonable request.}


\end{document}